\documentclass[11pt]{article}
\usepackage{amsfonts,amssymb, bm, graphicx,a4}
\usepackage{color}

\newtheorem{teo}{Theorem}[section]

\newtheorem{pro}{Proposition}[section]
\newtheorem{rema}{Remark}[section]

\definecolor{Red}{cmyk}{0,1,1,0}

\def\R{\mathbb{R}}
\def\0{\emptyset}

\begin{document}


\voffset=-1.5truecm\hsize=16.5truecm    \vsize=24.truecm
\baselineskip=14pt plus0.1pt minus0.1pt \parindent=12pt
\lineskip=4pt\lineskiplimit=0.1pt      \parskip=0.1pt plus1pt

\def\ds{\displaystyle}\def\st{\scriptstyle}\def\sst{\scriptscriptstyle}


\let\a=\alpha \let\b=\beta \let\chi=\chi \let\d=\delta \let\e=\varepsilon
\let\f=\varphi \let\g=\gamma \let\h=\eta    \let\k=\kappa \let\l=\lambda
\let\m=\mu \let\n=\nu \let\o=\omega    \let\p=\pi \let\ph=\varphi
\let\r=\rho \let\s=\sigma \let\t=\tau \let\th=\vartheta
\let\y=\upsilon \let\x=\xi \let\z=\zeta
\let\D=\Delta \let\F=\Phi \let\G=\Gamma \let\L=\Lambda \let\Th=\Theta
\let\O=\Omega \let\P=\Pi \let\Ps=\Psi \let\Si=\Sigma \let\X=\Xi
\let\Y=\Upsilon

\global\newcount\numsec\global\newcount\numfor
\gdef\profonditastruttura{\dp\strutbox}
\def\senondefinito#1{\expandafter\ifx\csname#1\endcsname\relax}
\def\SIA #1,#2,#3 {\senondefinito{#1#2}
\expandafter\xdef\csname #1#2\endcsname{#3} \else
\write16{???? il simbolo #2 e' gia' stato definito !!!!} \fi}
\def\etichetta(#1){(\veroparagrafo.\veraformula)
\SIA e,#1,(\veroparagrafo.\veraformula)
 \global\advance\numfor by 1
 \write16{ EQ \equ(#1) ha simbolo #1 }}
\def\etichettaa(#1){(A\veroparagrafo.\veraformula)
 \SIA e,#1,(A\veroparagrafo.\veraformula)
 \global\advance\numfor by 1\write16{ EQ \equ(#1) ha simbolo #1 }}
\def\BOZZA{\def\alato(##1){
 {\vtop to \profonditastruttura{\baselineskip
 \profonditastruttura\vss
 \rlap{\kern-\hsize\kern-1.2truecm{$\scriptstyle##1$}}}}}}
\def\alato(#1){}
\def\veroparagrafo{\number\numsec}\def\veraformula{\number\numfor}
\def\Eq(#1){\eqno{\etichetta(#1)\alato(#1)}}
\def\eq(#1){\etichetta(#1)\alato(#1)}
\def\Eqa(#1){\eqno{\etichettaa(#1)\alato(#1)}}
\def\eqa(#1){\etichettaa(#1)\alato(#1)}
\def\equ(#1){\senondefinito{e#1}$\clubsuit$#1\else\csname e#1\endcsname\fi}
\let\EQ=\Eq


\def\bb{\hbox{\vrule height0.4pt width0.4pt depth0.pt}}\newdimen\u
\def\pp #1 #2 {\rlap{\kern#1\u\raise#2\u\bb}}
\def\hhh{\rlap{\hbox{{\vrule height1.cm width0.pt depth1.cm}}}}
\def\ins #1 #2 #3 {\rlap{\kern#1\u\raise#2\u\hbox{$#3$}}}
\def\alt#1#2{\rlap{\hbox{{\vrule height#1truecm width0.pt depth#2truecm}}}}

\def\pallina{{\kern-0.4mm\raise-0.02cm\hbox{$\scriptscriptstyle\bullet$}}}
\def\palla{{\kern-0.6mm\raise-0.04cm\hbox{$\scriptstyle\bullet$}}}
\def\pallona{{\kern-0.7mm\raise-0.06cm\hbox{$\displaystyle\bullet$}}}


\def\data{\number\day/\ifcase\month\or gennaio \or febbraio \or marzo \or
aprile \or maggio \or giugno \or luglio \or agosto \or settembre
\or ottobre \or novembre \or dicembre \fi/\number\year}

\setbox200\hbox{$\scriptscriptstyle \data $}

\newcount\pgn \pgn=1
\def\foglio{\number\numsec:\number\pgn
\global\advance\pgn by 1}
\def\foglioa{a\number\numsec:\number\pgn
\global\advance\pgn by 1}



\def\sqr#1#2{{\vcenter{\vbox{\hrule height.#2pt
\hbox{\vrule width.#2pt height#1pt \kern#1pt
\vrule width.#2pt}\hrule height.#2pt}}}}
\def\square{\mathchoice\sqr34\sqr34\sqr{2.1}3\sqr{1.5}3}

\let\ciao=\bye \def\fiat{{}}
\def\pagina{{\vfill\eject}} \def\\{\noindent}
\def\bra#1{{\langle#1|}} \def\ket#1{{|#1\rangle}}
\def\media#1{{\langle#1\rangle}} \def\ie{\hbox{\it i.e.\ }}
\let\ii=\int \let\ig=\int \let\io=\infty

\let\dpr=\partial \def\V#1{\vec#1} \def\Dp{\V\dpr}
\def\oo{{\V\o}} \def\OO{{\V\O}} \def\uu{{\V\y}} \def\xxi{{\V \xi}}
\def\xx{{\V x}} \def\yy{{\V y}} \def\kk{{\V k}} \def\zz{{\V z}}
\def\rr{{\V r}} \def\zz{{\V z}} \def\ww{{\V w}}
\def\Fi{{\V \phi}}

\let\Rar=\Rightarrow
\let\rar=\rightarrow
\let\LRar=\Longrightarrow

\def\lh{\hat\l} \def\vh{\hat v}

\def\ul#1{\underline#1}
\def\ol#1{\overline#1}

\def\ps#1#2{\psi^{#1}_{#2}} \def\pst#1#2{\tilde\psi^{#1}_{#2}}
\def\pb{\bar\psi} \def\pt{\tilde\psi}

\def\E#1{{\cal E}_{(#1)}} \def\ET#1{{\cal E}^T_{(#1)}}
\def\LL{{\cal L}}\def\RR{{\cal R}}\def\SS{{\cal S}} \def\NN{{\cal N}}
\def\HH{{\cal H}}\def\GG{{\cal G}}\def\PP{{\cal P}} \def\AA{{\cal A}}
\def\BB{{\cal B}}\def\FF{{\cal F}}

\def\tende#1{\vtop{\ialign{##\crcr\rightarrowfill\crcr
              \noalign{\kern-1pt\nointerlineskip}
              \hskip3.pt${\scriptstyle #1}$\hskip3.pt\crcr}}}
\def\otto{{\kern-1.truept\leftarrow\kern-5.truept\to\kern-1.truept}}
\def\arm{{}}
\font\bigfnt=cmbx10 scaled\magstep1
\def\ind#1{\1_{\{#1\}}}
\def\TT{{\mathcal{T}}}
\def\1{\rlap{\mbox{\small\rm 1}}\kern.15em 1}
\def\Rd{\mathbb{R}^d}
\newcommand{\card}[1]{\left|#1\right|}

\BOZZA
\def\La{\Lambda}
\title{Cluster expansion  for continuous particle systems interacting  via  an  attractive
pair potential and subjected to high density  boundary conditions }
\author{
\\
Paula M. S. Fialho, Bernardo N. B. de Lima, Aldo Procacci\\
\\
\small{ Departamento de Matem\'atica UFMG}
\small{ 30161-970 - Belo Horizonte - MG
Brazil}}

\maketitle

\begin{abstract}
We propose a method based on cluster expansion to study
 the low activity/high temperature phase of a continuous particle system
 confined in a finite volume,
interacting through a stable and finite range pair potential
with negative minimum in presence of non free
boundary conditions.
\end{abstract}

\section{Introduction}

In the area of rigorous   statistical mechanics
from the very beginning  a great effort has been
spent in order to outline the possible influence of the boundary conditions on
systems confined in a finite (but possibly arbitrarily large)  volume.
It has been clear soon (see for instance \cite{Ru,Ga} and references therein) that in the regime in which
more phases may coexist the presence of suitable boundary conditions may force
the system in one of those phases. This has been rigourously established and put on firm ground for
a large class of bounded spin systems in a lattice
interacting via a finite range potential.
A classical example
is  the nearest-neighbor Ising model in two or more dimensions.
Similar results can be obtained for large classes of bounded spin systems with finite range interaction
for which a very robust and effective tool, the Pirogov Sinai theory, is available (see, e.g.  \cite{Za} and references therein).
The effect and influence of boundary conditions on spin systems which are unbounded or interact with  infinite range potential appears to be a more delicate issue
to be treated rigorously and results in the literature are quite rarer. Problems related to existence and
uniqueness of the infinite volume measure for unbounded spin-systems has been discussed e.g. in
\cite{LP}, \cite{COP} and
\cite{Pa}, while analyticity of  free energy and correlations for such systems subjected
to rather  general boundary conditions has been treated via cluster expansion in  \cite{APS} and \cite{PS}.

The situation is  even less clear as soon as one considers continuous systems formed by classical particles in $\Rd$ interacting
via a pair  potential (such as the Lennard-Jones potential, the Morse potential or even much simpler potentials, e.g. finite range). In this case the only phase which has been rigourously analyzed  is the
low densitity/high temperature phase and no proof on the existence of phase transitions has been furnished  nor a consistent and rigorous
treatment of such systems outside the low density/high temperature region has been provided with the sole exception of the result
obtained by Mazel, Lebowitz and Presutti in 1999 \cite{LMP}.
\newpage

The relation between the boundary condition and the macroscopical behavior of continuous classical particle systems could in principle
be studied  rigorously at least  in the low density/high temperature regime where the powerful tool
given by the cluster expansion is available.
This problem is incidentally mentioned by the classical texts on rigorous statistical mechanics (see e.g. \cite{Ru, Ga}),
where however the computations related to the
analyticity of the pressure of the gas in the low density/high temperature regime are  always performed assuming  free boundary conditions.
Although  it is widely believed that the macroscopic behavior
of continuous particle systems in the region of parameters $\l$ and $\b$ where the cluster expansion converges
is not affected by (reasonably well behaved) boundary conditions,
we are not aware of any rigorous result about this issue as soon as one considers pair potentials with a negative (attractive) tail.
In particular, the  independence of  convergence radius of the cluster expansion from (reasonable) boundary conditions can be estabilshed only
by assuming that the pair potential $v$ is non-negative,
 in addition to stable  and regular (see e.g. the remark at the end of pag. 3 in \cite{Sp}). Of course the  assumption $v\ge0$
rules out nearly all   physically relevant pair potential which  are usually attractive (and thus negative) at large distances.
In this respect it is symptomatic to observe that all results
on the dynamics of continuous particle systems which use directly or indirectly the convergence of the cluster expansion
treat only systems interacting via positive potentials (see e.g. \cite{BCC,BCDP, Ja, K,KKM, KL,Ma,Sp,Wu} ).

Rigorous result on the influence of boundary conditions in classical particle systems  has been only of very general nature, such as
well-definiteness and existence problems of the infinite volume measure for such systems (see e.g. \cite{KPR} and references therein).
In particular, the problem of uniqueness of the infinite volume measure (in the high temperature/low density region) has  been discussed
in \cite{Le}, \cite{PZ},  \cite{KKO} and references therein.
To our knowledge nothing has been published so far about how the analyticity region of the pressure
may be affected by the boundary conditions in continuous particle systems interacting via non purely repulsive potentials.

In this paper we consider  a system of
continuous classical particles in $d$ dimensions  confined in a
finite volume $\L$ in the grand canonical ensemble at fixed inverse temperature $\beta$ and fixed fugacity $\lambda$.
We assume that these particles interact via a stable  pair potential $v$ which, for simplicity,  we impose to be finite range.
On the other hand,  we allow $v$ to have a negative tail. In other words
the stability constant of $v$  may be  strictly positive. This system would have a fixed mean density $\r^\0_\L(\l,\b)$ when submitted  to  free boundary conditions.  We then fix a boundary
configuration $\omega$ outside $\L$ (i.e. in $\mathbb{R}^d\setminus \L$) allowing
a density $\rho_\o$  which  may be much larger than $\r^\0_\L(\b,\l)$
but has to be uniformly bounded.
Note that we allow $\r_\o$ to be arbitrary so that we are actually allowing boundary condition with arbitrarily large (but bounded)
densities. With these assumptions we show that the Mayer series of
the pressure of the system in presence of the boundary condition $\o$ can be written as the sum of two terms. The first
series, the bulk term,   has a radius of analyticity in the activity $\lambda$ that coincides with
the free boundary condition convergence radius. The second series, the boundary term,  has  an $\o$-dependent radius of analyticity decreasing
exponentially with $\rho_\o$, but
it tends to zero as $\Lambda$ goes to infinity. Moreover, we show that the bulk term of the finite volume pressure in presence of
boundary conditions $\o$ tends to the  Mayer series of the pressure calculated with free boundary conditions.

\section{The model and results}\label{sec2}
\numsec=2\numfor=1

We consider a system of classical continuous particles confined in a bounded compact region $\L$ of $\mathbb{R}^d$
interacting via a translational invariant  pair potential $v$. We will suppose hereafter that $\L$ is a cube of size $2L$ centered at the origin
and $\lim{\L\to \infty}$ means simply that $L\to \infty$. We will denote by $|\L|=(2L)^d$ the volume of $\L$ and in general if $U$ is a compact in $\mathbb{R}^d$ we denote by $|U|$ its volume. We denote by $x_i\in \mathbb{R}^d$
the position vector of the $i^{\rm th}$ particle of the system and by $\|x_i\|$ its Euclidean norm.


We will further suppose that our system is  subjected  to a  boundary condition $\o$ which  is typically a locally finite
countable set of points in $\mathbb{R}^d$ (not necessarily distinct), representing the positions
of a set of fixed particles in $\mathbb{R}^d$. Namely, $\o$ is a set
such that for any compact subset $C\subset \mathbb{R}^d$,  $\#(\o\cap
C)<+\infty$ (here $\#(\o\cap C)$ is the cardinality of the set
$\o\cap C$).
As usual, we will suppose that each particle inside $\L$, say at position $x\in \L$,
feels the effect of boundary condition
$\o$ through the potential energy  generated by the particles of the configuration $\o$ which are in $\L^c=\mathbb{R}^d\setminus\L$.
We are interested in  studying  the behavior of the systems in the limit $\L\to\infty$ for a fixed boundary condition
$\o$ and how eventually this limit
may be  influenced by $\o$, having in mind that as the volume $\L$  invades $\mathbb{R}^d$ the fixed particles of $\o$ entering in $\L$ are disregarded
and
only those boundary particles outside $\L$  influence particles inside $\L$.

\vskip.3cm
\\\underline{\it Assumptions on the pair potential}
\vskip.1cm
The pair potential $v$ is supposed to be  a translational invariant, symmetric and   Lebesgue measurable function. Therefore it is
completely defined by a function $v(x)$ in $\R^d$ with values in $\mathbb{R}\cup\{+\infty\}$ such that
$v(x)=v(-x)$ for all $x\in \mathbb{R}^d$.  We further assume that
\vskip.2cm
\\
\begin{itemize}
\item[\rm (ii)] {\it $v$ is {\it finite range}}:
\\ there exists $R>0$ such that
$$
v(x)=0  ~~~~~{\rm for ~all}~~~\|x\|\ge R. \Eq(2.7)
$$
\vskip.2cm
\item[\rm (i)]{\it $v$ is stable}:
\\namely,  $v$ is such that for some constant $C\ge 0$,
for all $n\in \mathbb{N}$ and for all $x_1,\dots,x_n\in \mathbb{R}^d$
$$
\sum_{1\le i<j\le n} v(x_i-x_j)\ge -C n.\Eq(2.6)
$$

\vskip.2cm
\end{itemize}

The optimal constant $C$ in \equ(2.6)  is called the {\it stability constant} of the potential $v$ and will be denote with the symbol ${\mathcal{B}}_v$. Then
$$
{\mathcal{B}}_v= \sup_{n\ge 2\atop (x_i,\dots,x_n)\in \mathbb{R}^{dn}}\Big\{-{1\over n}\sum_{1\le i<j\le n}v(x_i-x_j)\Big\}.
$$
Note that if ${\mathcal{B}}_v>0$ then there are  points $x\in \Rd$ such that $v(x)<0$. Let us denote by $v^-$ the  negative part of $v$, namely, for $r\in [0,+\infty)$,
$$
v^-(r)=\max\{0, - v(r)\}.
$$

By \equ(2.6), the potential  $v$  is bounded below by $-2{\mathcal{B}}_v$ and hence $v^-$  is bounded above by $2{\mathcal{B}}_v$.
Therefore, by \equ(2.7), we have that
$$
\int_{\mathbb{R}^d} v^-(x)dx \le 2{\mathcal{B}}_vV_d(R),\Eq(deF)
$$
where $V_d(R)$ is the volume of the $d$-dimensional ball of radius $R$.

We will suppose hereafter
that $\R^d$ is partitioned in elementary cubes $\D$ of suitable size $\d>0$.
Along the paper we will  denote by $\R^d_\d$ the set of all these cubes and, given $x\in \Rd$, we will denote by $\D(x)$ the cube of $\Rd_\d$ to which $x$ belongs. Moreover, for sake of simplicity  we assume that $\L$ is so chosen in such a way
$\L$ and $\L^c$ are both the union of elementary cubes in  $\R^d_\d$ (in other words, for any $\D\in \R^d_\d$, either $\D\subset \L$ or $\D\subset \L^c$). We denote by $\L_\d$ (respectively $\L^c_\d$) the set of elementary cubes whose union is $\L$ (respectively $\L^c$) and of course by construction $\L_\d\cup\L^c_\d=\R^d_\d$.
Given $\D\in  \Rd_\d$ and $x\in \Rd$ we let  ${\rm dist}(\D,x)=\inf_{y\in \D}\|x-y\|$  and  we will suppose that $\d$ is chosen suitably small in such a way  so that   for any $x\in \Rd$
$$
\d^d\sum_{\D\in \Rd_\d\atop {\rm dist}(\D,x)\le R}1\le 2 V_d(R).\Eq(tusf)
$$

\vskip.3cm
\\\underline{\it Assumptions on   boundary conditions}

\vskip.1cm

\\Given  the partition of $\Rd$ in elementary cubes $\D$ of  size $\d>0$ described above and given   a locally finite boundary condition $\o$, we  define the density of $\o$ as the function
\begin{eqnarray}
 \nonumber \r_\d^\o &:& \Rd\to [0, +\infty) \\
 \nonumber   && x\mapsto \r_\d^\o(x)
\end{eqnarray}

 \\with
$$\r_\d^\o(x)=\frac{\#(\o\cap \D(x))}{\d^{-d}},$$
then, by definition, $\r_\d^\o(y)$ is constant for all $y\in \D(x)$. Since $\o$ is locally finite, $\r_\d^\o(x)$ is everywhere finite.
\vskip.2cm

Our assumption on the set of allowed boundary conditions 
is as follows.
\begin{itemize}
\item[\rm (iii)]{\it $\o$ is admissible:} namely,  there exists a finite positive number $\r_\o$ such that,
for all elementary cubes $\D\subset \mathbb{R}_\d^d$,
$$
\sup_{\D\subset \mathbb{R}^d_\d}{\#(\o\cap \D)\over |\D|}\le \r_\o.
$$
\end{itemize}

We call   $\O_\r$ the space of all locally finite configurations of particles
in $\mathbb{R}^d$ with maximal density $\r$ and we set $\O^*=\cup_{\r\ge 0}\O_\r$.
Note that the free boundary condition $\o=\0$ is obviously in $\O_\r$, for all $\r>0$.
\vskip.2cm

Let $\partial \L$ denotes the boundary of $\L$ and let us define,  for $x\in \L$ fixed,
$$
d_x^\L= {\rm dist} (x, \partial \L)=\inf_{y\in \partial\L}\|x-y\|.
$$
For a fixed volume $\L$ and a fixed boundary condition $\o$, let us define  the function $w^\o_\L: \L\to \R$ as follows
$$
w^\o_\L(x)=  \sum_{y \in \o\cap \L^c} v(x-y).\Eq(fa)
$$
This function represents the potential energy felt by a particle sitting in the point $x\in \L$ due to the
fixed particles of the boundary condition $\o$ sitting in points outside $\L$.
Note that by the assumptions {\rm (ii)}  on the pair potential and {\rm (iii)} on the admissible boundary conditions we have that
$w^\o_\L(x)$ is different from zero only in the frame inside $\L$ constituted by the points at distance less than $R$
to the boundary $\partial \L$.

The partition function of the system in the grand canonical ensemble at fixed inverse temperature $\b>0$ and fixed fugacity $\l>0$
is given by
$$
\Xi^\o_{\La}(\l,\b)=\sum_{n=0}^{\infty}{\l^{n}\over n!}
\int_{\L^n} dx_1\dots dx_n e^{-\b\left[\sum\limits_{1\le i< j\le n}v(x_i-x_j)+\sum\limits_{i=1}^nw^\o_\L(x_i)\right]}.\Eq(1.1)
$$
It is easy to check that with our assumptions on $v$ and $\o$ the power series in the l.h.s. of \equ (1.1) is absolutely convergent for all $\l\in \mathbb{C}$.
Indeed the bulk factor $\sum_{1\le i< j\le n}v(x_i-x_j)$ in the exponent of the integrand in the r.h.s. of \equ(1.1) is greater  than $-Cn$  by the stability condition
\equ(2.6) and the boundary factor  $\sum_{i=1}^nw^\o_\L(x_i)$ in the same exponent is also bounded from below by a negative  constant times $n$, as shown in the following proposition.
\vskip.2cm
\begin{pro}\label{pa}
If $\o\in \O^*$, then for all $x\in \L$ we have
$$
w^\o_\L(x)  ~~   \cases{~= 0~ & ~~~if ~$d_x^\L\ge R$ \cr\cr ~\ge~ -\k\r_\o  &~~ if ~$d_x^\L<R$ }\Eq(impo)
$$
where $\k=4{\mathcal{B}}_vV_d(R)$.
\end{pro}
\vskip.2cm
{\bf Proof.}
If $d_x^\L\ge R$,  then $ v(x-y)=0$ for all $y\in \L^c$ and therefore, by definition \equ(fa),  $w^\o_\L(x)=0$. On the other hand, if $d_x^\L< R$ then
$$
w^\o_\L(x)=  \sum_{y \in \o\cap\L^c} v(x-y)\ge-\sum_{y \in \o\cap\L^c} v^-(x-y)=
-\sum_{\D\in \L^c_\d} \sum_{y \in \o\cap \D}  v^-(x-y)\ge
$$
$$
\ge
-\sum_{\D\in\L^c_\d \atop
{\rm dist}(\D,x)< R}2{\mathcal{B}}_v\sum_{y\in \o\cap \D} 1 \ge -2{\mathcal{B}}_v
\r_\o|\D|\sum_{\D\in\L^c_\d \atop
{\rm dist}(\D,x)< R}1\ge -4{\mathcal{B}}_v V_d(R)
\r_\o
$$
where in the last inequality we have used \equ(tusf).

~~~~~~~~~~~~~~~~~~~~~~~~~~~~~~~~~~~~~~~~~~~~~~~~~~~~~~~~~~~~~~~~~~~~~~~~~~~~~~~~~~~~~~~~~~~~~~~~~~~~~~~~~~~~~~~~~~~~~~~~~~~~~~~~~~~~~~~~~~~~~~~~~~~~~~~~~~~~$\Box$

\vskip.3cm

The finite volume pressure of the system is  given by
$$
\b p_\L^\o(\l,\b)={1\over |\L|}\log \Xi^\o_{\La}(\l,\b)\Eq(pressureb)
$$
and the thermodynamic limit of the finite volume pressure (if it exists)  is
$$
\b p^\o(\l,\b)=\lim_{\L\to\infty}{1\over |\L|}\log \Xi^\o_{\La}(\l,\b).\Eq(prlim)
$$

The existence of the limit \equ(prlim) when $\o=\0$ (i.e in presence of free boundary conditions)
 is a well established fact since the sixties. Namely, it has been proved (see \cite{Ru, Ru70} and references therein) that $\b p^\0(\l,\b)$
 exists    and  it is  continuous as a function of $\l$ and $\b$ in the whole physics domain $(\l,\b)\in [0,+\infty)\times [0,+\infty)$, as soon as particles interact
via a superstable and regular pair potential. We recall that  a  pair potential $v$ is {\it superstable} if it can be written as $v=v_1+v_2$,
where both  $v_1$ and $v_2$ are translational invariant, symmetric and Lebesgue measurable real valued functions in $\Rd$, $v_1$ is stable
and $v_2$ is non-negative and such that there exists a positive constant $a$  such that $v_2(x)\ge v_2(a)>0$ for all $|x|<a$. We also recall that
$v$ is {\it regular} if the function $f(x)=e^{-v(x)}-1$ is absolutely summable in $\Rd$.

Later,   Georgii \cite{Ge,Ge2}  studied
the limit \equ(prlim) when $\o\neq \0$ and he showed that it exists and   it is equal to $\b p^\0(\l,\b)$ as soon as $\o$ varies in a large class of allowed boundary conditions which Georgii called ``tempered boundary conditions $\o$'' (see (2.24) in \cite{Ge}  or (2.6) in \cite{Ge2}) provided that the pair potential $v$, beyond superstable   and regular,  has a hard-core or diverges in a non summable way at short distances. According to Georgii a boundary condition $\o$ is {\it tempered}  if, for some finite positive constant $t$,  $\limsup_{\La\to \infty} |\L|^{-1}\sum_{\D_\d\in \L_\d}[\#(\o\cap \D_\d)]^2\le t$, where $\L_\d$ is a collection of cubes $\D_\d$ of fixed size $\d>0$ forming a partition of $\L$. It is also worth to mention that very recently it has been  shown  \cite{PY2} that
the limit \equ(prlim) exists and it is equal to $\b p^\0(\l,\b)$ assuming just  stability and regularity of $v$
and  considering boundary conditions $\o$ with a density $\r_\o$ possibly growing to infinity with the distance from the origin at a rate which depends on decay of $v$ at large distances.

While the superstability condition is required by \cite{Ge}, \cite{Ge2}, \cite{PY} and \cite{PY2} to show the existence of the limit \equ(prlim) for all positive values of $\l$ and $\b$, in this paper we just need $v$ to be stable once we are only interested in the region of parameters $\l$ and $\b$ such that  $|\l|<\RR(\b)$, where $\RR(\b)$ is the convergence radius of the cluster expansion.

Concerning specifically the analyticity of  $\b p^\o(\l,\b)$ at low densities,  it has  been shown in the sixties (see e.g. \cite{Ru} and references therein) that when $\o=\0$,
both the finite volume pressure $\b p_\L^\0(\l,\b)$ and the infinite volume pressure $\b p^\0(\l,\b)$ can be written in terms
of  power series in $\l$ (the finite volume  Mayer series and the infinite volume Mayer series) which are analytic
 for all complex $\l$ in a disc around $\l=0$ as far as the potential $v$ is stable and regular. The  radius $\RR(\b)$ of the analyticity disc is  uniformly bounded below by a constant depending only on the temperature
 and the potential. The best constant for the lower bound of  $\RR(\b)$ for continuous particle systems
interacting via stable and regular pair potentials has been given recently by Procacci and Yuhjtman in \cite{PY} where it is proved that
$$
\RR(\b) \ge   {1\over e^{\b B+1} C_v(\b)} \Eq(py)
$$
with
$$
C_v(\b)= \int_{\mathbb{R}^d} d x (1-e^{-\b |v(x)|}).\Eq(cvbeta)
$$

The technique used to write $\log \Xi^\0_{\La}(\l,\b)$ is terms of a convergent series can be naively extended also when $\o\neq \0$, but  in this  case  the convergence radius depends on $\o$ and in general  tends to shrink to zero if $\r_\o\to \infty$ (unless the potential $v$ is non-negative, \cite{Sp}). In this paper, we will derive   a non-naive expansion of $|\L|^{-1}\log \Xi^\o_{\La}(\l,\b)$ in terms of powers  of $\l$ whose coefficients
depend on the inverse temperature $\b$, the volume $\L$ and the boundary condition $\o$ (the so called Mayer series)
and we will analise the behavior of    this series  when $\L$ goes to infinity  and  $\o$ varies in $\O^*$. Our main result can be summarized by the following theorem.
\begin{teo}\label{main}
Let $v$ satisfies assumption {\rm (i)} and {\rm (ii)} and let $\o\in \O^*$.
Let  $\mathcal{D}^\0$ be the closed disc in the complex plane
$$
\mathcal{D}^\0=\Big\{\l\in \mathbb{C}:~ |\l| \le {1\over e^{\b {\mathcal{B}}_v+1}C_v(\b)}\Big\}\Eq(dv)
$$
and let  $\mathcal{D}^\o$ be the closed disc in the complex plane
$$
\mathcal{D}^\o=\Big\{\l\in \mathbb{C}:~ |\l| \le {1\over e^{\k\b\r_\o}e^{\b {\mathcal{B}}_v+1}C_v(\b)}\Big\}.\Eq(dnv)
$$

\\Then, for all $\L$ and all $\o\in \O^*$ the finite volume pressure $p^\o_\L(\b,\l)$ of the system is such that
$$
\b p^\o_\L(\b,\l)=  \h^\o_\L(\l,\b)+ \x^\o_\L(\l,\b),
$$
where
\begin{itemize}
\item[a)] $\h^\o_\L(\l,\b)$ is analytic in $\l$   in the disk $\mathcal{D}^\0$ where, uniformly in $\L$, admits the bound
$$
|\h^\o_\L(\l,\b)|\le  (8/7)e^{\b {\mathcal{B}}_v+1}|\l|.
$$
\item[b)] $\x^\o_\L(\l,\b)$ is
analytic in $\l$   in the disk $\mathcal{D}^\o$ where, uniformly in $\L$, admits the bound
$$
|\x^\o_\L(\l,\b)|\le |\l| e^{\b\k\r_\o} e^{\b {\mathcal{B}}_v+1} g(\L)
$$
for some $g(\L)$ such that
$$
\lim_{\L\to \infty}g(\L)=0
$$
and thus
$$
\lim_{\L\to \infty}\x^\o_\L(\l,\b)=0.
$$
\item[c)] For all  $\l\in \mathcal{D}^\0$ it holds
$$
\lim_{\L\to \infty} [\h^\o_\L(\l,\b)- p^\0_\L(\l,\b)]=0.
$$

\end{itemize}

\end{teo}


\section{Proof of Theorem \ref{main}}
\subsection{Mayer expansion}
\numfor=1\numsec=3

\\We  start by rewriting the partition function \equ(1.1) of the system subjected to the  boundary condition $\o\in \O^*$ as follows

$$
\Xi^\o_{\La}(\l,\b)=\sum_{n=0}^{\infty}{\l^n\over n!}
\int_\L dx_1\dots \int_\L dx_n e^{-\b\sum_{1\le i< j\le n}v(x_i-x_j)}f^\o_\L(x_1)\dots f^\o_\L(x_n),
\Eq(1.2)
$$
where
$$
f^\o_{\L}(x)= e^{- \b w^\o_\L(x)} \Eq(hx)
$$
with $w^\o_\L(x)$ defined in \equ(fa).

\begin{rema}\label{f}
  By Proposition \ref{pa} we have that
$$
f^\o_{\L}(x)  ~~   \cases{~= 1~ & ~~~if ~$d_x^\L\ge R$ \cr\cr ~\le~ e^{\b\k\r_\o}  &~~ if ~$d_x^\L<R$ }.
$$
\end{rema}

\vskip.2cm


It is then a standard, but not trivial (see \cite{Mc}), task to show 
that the logarithm of $\Xi^\o_{\La}(\l,\b)$ can be written  as follows

$$
\log\Xi^\o_{\La}(\l,\b)=\sum_{n=1}^{\infty}{\l^n\over n!}
\int_\L dx_1\dots \int_\L dx_n \Phi^{T}(x_1,\dots, x_n)f^\o_\L(x_1)\dots f^\o_\L(x_n)
\Eq(1.2.b)
$$
with
$$
\Phi^T(x_1,\dots,x_n)~=~ \cases{\sum\limits_{g\in G_{n}}
\prod\limits_{\{i,j\}\in E_g}\left[  e^{ -\b v(x_i -x_j)} -1\right] &if
$n\ge 2$\cr \cr 1 &if $n~=~ 1$}
\Eq(urse)$$
where  $G_n$ is the set of all connected graphs $g$ with vertex set
$[n]\doteq\{1,2,\dots, n\}$ and edge set  $E_g$.

The Mayer series of the (finite volume) pressure in presence of non free boundary conditions $\o$ is defined as the power series \equ(1.2.b) divided by $|\L|$, namely,
$$
\b p^\o_\L(\b,\l)\doteq{1\over |\L|}\log\Xi^\o_{\La}(\l,\b)=\sum_{n=1}^\infty c_n^\o(\b,\L)\l^n, \Eq(mayer)
$$
where
$$
c_n^\o(\b,\L)={1\over |\L|} {1\over n!} \int_\L dx_1\dots \int_\L dx_n \Phi^{T}(x_1,\dots, x_n)f^\o_\L(x_1)\dots f^\o_\L(x_n).
$$
Note that we can also write
$$
\log\Xi^\o_{\La}(\l,\b)= \l\int_\L dx_0 f^\o_\L(x_0)\Pi^\o_{x_0,\L}(\b,\l),\Eq(1.2.c)
$$
where
$$\Pi^\o_{x_0,\L}(\b,\l)=
\sum_{n=0}^{\infty} c^\o_n(x_0,\b,\L){\l^n}
\Eq(1.3)
$$
and
$$
c^\o_n(x_0,\b,\L)= {1\over (n+1)!}\int_\L dx_1\dots \int_\L dx_n \Phi^{T}(x_0,x_1,\dots, x_n)
 f^\o_\L(x_1)\dots f^\o_\L(x_n),\Eq(cex)
$$
with
$$
\Phi^T(x_0,x_1,\dots,x_n)~=~ \cases{\sum\limits_{g\in G^0_{n}}
\prod\limits_{\{i,j\}\in E_g}\left[  e^{ -\b v(x_i -x_j)} -1\right] &if
$n\ge 1$\cr \cr 1 &if $n~=~ 0$}
\Eq(urseb)$$
where  $G^0_n$ denotes  now the set of all connected graphs $g$ with vertex set
$[n]_0\doteq\{0,1,2,\dots, n\}$ and edge set  $E_g$. We agree that $c^\o_n(x_0,\b,\L)=1$ if $n=0$.

Using the above notations, the pressure of the system  at finite volume can  also be written as
$$
\b p^\o_\L(\b,\l)~=~{\l\over |\L|}\int_\L dx f^\o_\L(x_0)\Pi^\o_{x_0,\L}(\b,\l),\Eq(pressure)
$$
which is an useful expression for the computations  developed ahead. We conclude this section by proving the following inequality concerning the absolute value of
the coefficients $c^\o_n(x_0,\b,\L)$ definied above.
\vskip.2cm
\begin{pro}\label{Cgene}
For any $x\in \L$ and any $\o\in \O^*$, it holds that
$$
|c^\o_n(x_0,\b,\L)|\le  {e^{\b  \k \r_\o n}}{(n+1)^{n-1}\over (n+1)!}  e^{\b {\mathcal{B}}_v(n+1)} [C_v(\b)]^n.\Eq(cogen)
$$
with $C_v(\b)$ defined in \equ(cvbeta).
\end{pro}
{\bf Proof}. We first recall  that the potential $v$  is  stable with  stability constant ${\mathcal{B}}_v$, therefore we can use the bound proved in \cite{PY}  (see there Proposition 1), namely,
$$
|\Phi^{T}(x_0,x_1,\dots, x_n)|\le  e^{\b {\mathcal{B}}_v(n+1)}\sum_{\tau \in T^0_{n}} \prod_{\{i,j\} \in E_\tau} (1-e^{-\b| V(x_i-x_j)|}), \Eq(PY)
$$
where $ T^0_{n}$ is the set of trees with vertex set $\{0,1,2,\dots, n\}$. Moreover, for any $x\in \L$, by Remark \ref{f} we have that $ f^{\o}_{\L}(x)\le e^{\k\b\r_\o}$. Thus
\begin{eqnarray}
\nonumber |c^\o_n(x_0,\b,\L)|&\le& {1\over (n+1)!}\int_\L dx_1\dots \int_\L dx_n |\Phi^{T}(x_0,x_1,\dots, x_n)| f^\o_\L(x_1)\dots f^\o_\L(x_n)\\
\nonumber &\le& e^{\b \k \r_\o n}\left[{1\over (n+1)!}\int_\L dx_1\dots \int_\L dx_n  e^{\b {\mathcal{B}}_v(n+1)}\sum_{\tau \in T^0_{n}} \prod_{\{i,j\} \in E_\tau} (1-e^{-\b| V(x_i-x_j)|})\right]\\
\nonumber &\le& e^{\b \k \r_\o n}\left[{ e^{\b {\mathcal{B}}_v(n+1)}\over (n+1)!}\sum_{\tau \in T^0_{n}}\int_\L dx_1\dots \int_\L dx_n  \prod_{\{i,j\} \in E_\tau} (1-e^{-\b| V(x_i-x_j)|})\right].
\end{eqnarray}
Now, for any $n\in \mathbb{N}$ and $\t\in T^0_{n}$ we have  (see e.g. Lemma 3 in \cite{PY})
$$
 \int_\L dx_1\dots \int_\L dx_n \prod_{\{i,j\} \in E_\tau} (1-e^{-\b| V(x_i-x_j)|})\le [C_v(\b)]^n.
 \Eq(2)
$$
Therefore,
\begin{eqnarray}
 \nonumber |c^\o_n(x_0,\b,\L)|&\le & e^{\b \k \r_\o n}{ e^{\b {\mathcal{B}}_v(n+1)}\over (n+1)!}[C_v(\b)]^n\sum_{\tau \in T^0_{n}}1 \\
\nonumber &\le& e^{\b \k \r_\o n}{ e^{\b {\mathcal{B}}_v(n+1)}\over (n+1)!}[C_v(\b)]^n (n+1)^{n-1}
\end{eqnarray}
where in the last line we have used the Cayley formula (see \cite{Ca} ), i.e. $|T^0_n|= (n+1)^{n-1}$.

~~~~~~~~~~~~~~~~~~~~~~~~~~~~~~~~~~~~~~~~~~~~~~~~~~~~~~~~~~~~~~~~~~~~~~~~~~~~~~~~~~~~~~~~~~~~~~~~~~~~~~~~~~~~~~~~~~~~~~~~~~~~~~~~$\Box$

We stress that bound \equ(cogen) is very crude and it may be quite strongly improved depending on the distance of  the point $x_0$ from the border
of $\L$. We will analyze in some more detail the behaviour of the coefficients $c^\o_n(x_0,\b,\L)$ in the next section.

\subsection{On the behavior of  $c^\o_n(x_0,\b,\L)$ }\label{s3.2}
Recall that we are supposing  $\L$ to be  a cube of size $2L$ centered at the origin.
Let us choose a monotonic  increasing continuous function
$h(L)$ such that   $\lim_{L\to \infty} h(L)=\infty$,  $\lim_{L\to \infty} h(L)/L=0$
and define
$$
\L_h=\{x\in \L: d_x^\L>h(L)\}
$$
and
$$
\L^{*}_h=\L\setminus \L_h
$$
so that $\L_h$ is a cube centered at the origin with size $2(L-h(L))$ fully contained  in $\L$ and  clearly
$$
\lim_{\L\to \infty} {|\L_h|\over |\L|}=1 \Eq(uno)
$$
and
$$
\lim_{\L\to \infty} {|\L^*_h|\over |\L|}=0. \Eq(due)
$$

\begin{rema}
  Observe that $\sqrt{L}$ is an example of a function that satisfies the proprieties described above for the function $h(L)$. However, while the properties \equ(uno) and \equ(due) are essential for our task, the function rule of $h(L)$ does not play an important role in the calculations ahead.
\end{rema}

Let us now choose $L$ large enough in such a way that $h(L)>R$, so that
$$
n_{h(L)} \doteq \left\lfloor{h(L)\over R}-1\right\rfloor\Eq(nl)
$$
is greater that or equal to  one. Observe that
$$
\lim_{\L\to\infty}n_{h(L)}=+\infty. \Eq(limn)
$$

\vskip.2cm
\begin{teo}\label{t1}
Let $x_0\in \L_h$ and let  $\o\in \O^*$. Then, for all $n\le n_{h(L)}$, we have that
$$
c^\o_n(x_0,\b,\L)= c^\0_n(x_0,\b,\L).  \Eq(inside)
$$
Moreover, for all $n\in \mathbb{N}\cup\{0\}$ the following bound holds
$$
|c^\0_n(x_0,\b,\L)|\le {(n+1)^{n-1}\over (n+1)!}  e^{\b {\mathcal{B}}_v(n+1)} [C_v(\b)]^n. \Eq(c0n)
$$
\end{teo}
\vskip.2cm
\\{\bf Proof}.
Let us start by  proving identity \equ(inside). We recall  the definition \equ(urseb) of $\Phi^{T}(x_0,x_1,\dots, x_n)$. If $g$ is any connected graph  with vertex set
$[n]_0$,
as $v$ is finite range with $v(x)=0$ if $\|x\|\ge R$, we have that
$$
\prod\limits_{\{i,j\}\in E_g}\left[  e^{ -\b v(x_i -x_j)} -1\right]=0 \Eq(g0)
$$
whenever for some $i\in [n]$, $\|x_0-x_i\|\ge nR$.  Indeed, given $g\in G^0_n$, suppose that there exists a vertex  $i\in [n]$ of $g$ such that $\|x_0-x_i\|\ge nR$. Then there exists a path $p_{i,0}=\{0,i_1,i_2,\dots, i_{k-1},i_k\equiv i\}$ contained in $g$ connecting $0$ to $i$, once $g$ is connected. Let $E_{p_{i,0}}$ be the edge set of such a path. Therefore
$$
\prod\limits_{\{i,j\}\in E_g}\left[  e^{ -\b v(x_i -x_j)} -1\right]=\prod\limits_{\{i,j\}\in E_g\setminus E_{p_{i,0}}}\left[  e^{ -\b v(x_i -x_j)} -1\right]
\prod\limits_{s=1}^k\left[  e^{ -\b v(x_{s-1} -x_s)} -1\right].
$$
Observe that in any case $k\le n$, the hypothesis that $\|x_0-x_i\|\ge nR$ implies that at least for one $s\in [k]$ we have that $\|x_{s-1}-x_s\|\ge R$ and thus $v(x_{s-1} -x_s)=0$, so that $e^{ -\b v(x_{s-1} -x_s)} -1=0$. In conclusion, if $\|x_0-x_i\|\ge nR$, then
$$
\prod\limits_{s=1}^k\left[  e^{ -\b v(x_{s-1} -x_s)} -1\right]=0
$$
and thus \equ(g0) follows.

The discussion above  immediately implies that $\Phi^{T}(x_0,x_1,\dots, x_n)=0$ if there exists  $i\in [n]$ such that $\|x_i-x_0\|>nR$ and thus we can rewrite $c^\o_n(x_0,\b,\L)$ as follows
$$
c^\o_n(x_0,\b,\L)= {1\over (n+1)!}\int_{x_1\in \L\atop \|x_0-x_1\|\le nR} dx_1\dots \int_{x_n\in \L\atop \|x_0-x_n\|\le nR} dx_n \Phi^{T}(x_0,x_1,\dots, x_n) f^\o_\L(x_1)\dots f^\o_\L(x_n).\Eq(cexb)
$$

Let us now suppose that $n\le  n_{h(L)}$, i.e.,
$$
n\le {h(L)\over R}-1,
$$
whence, as by hypotheses $x_0\in \L_h$, then $d_{x_0}^\L\ge h(L)$ and we have
$$
d_{x_0}^\L\ge (n+1)R. \Eq(dist0)
$$
Moreover, by the triangular  inequality,
$$
d_{x_i}^\L\ge d_{x_0}^\L-\|x_i-x_0\|,
$$
hence
$$
d_{x_i}^\L\ge (n+1)R- \|x_i-x_0\|\ge (n+1)R- nR\ge R
$$
where in the intermediate inequality we used \equ(dist0) and in the last inequality we used that any $n$-uple $(x_1,\dots,x_n)$ contributing to the integral of the r.h.s. of \equ(cexb) is such that,  for any $i\in [n]$, $\|x_i-x_0\|\le nR$.

In conclusion we have shown  that if $n\le  n_{h(L)}$, then  for  any $n$-uple $(x_1,\dots,x_n)\in \L^n$ such that $\|x_0-x_i\|\le nR$ for all $i\in [n]$, it holds that $d_{x_i}^\L\ge R$. Recalling the Remark \ref{f}, we have that  $f^\o_\L(x_i)=1$ for all $i\in [n]$ in formula \equ(cex) when $n\le  n_{h(L)}$. Therefore, we have the identity
$$
c^\o_n(x_0,\b,\L)= {1\over (n+1)!}\int_\L dx_1\dots \int_\L dx_n \Phi^{T}(x_0,x_1,\dots, x_n) =  c^\0_n(x_0,\b,\L)
$$
for all  $x_0\in \L_h$  and for all  $n\le  n_{h(L)}$. Namely, we have proved the statement \equ(inside).

Let us now prove bound \equ(c0n).
Recalling that the potential $v$  is  stable with  stability constant ${\mathcal{B}}_v$, we can once again use the bound \equ(PY), namely,
$$
|\Phi^{T}(x_0,x_1,\dots, x_n)|\le  e^{\b {\mathcal{B}}_v(n+1)}\sum_{\tau \in T^0_{n}} \prod_{\{i,j\} \in E_\tau} (1-e^{-\b| V(x_i-x_j)|}),
$$
where $ T^0_{n}$ is the set of trees with vertex set $\{0,1,2,\dots, n\}$. Thus
$$
\int_\L dx_1\dots \int_\L dx_n |\Phi^{T}(x_0,x_1,\dots, x_n)|\le
e^{\b {\mathcal{B}}_v(n+1)}\sum_{\tau \in T^0_{n}} \int_\L dx_1\dots \int_\L dx_n \prod_{\{i,j\} \in E_\tau} (1-e^{-\b| V(x_i-x_j)|}).
\Eq(1)
$$
Now, for any $n\in \mathbb{N}$ and $\t\in T^0_{n}$ we can use again inequality \equ(2)
and hence  we have the upper bound
\begin{eqnarray}
\nonumber  c^\0_n(x_0,\b,\L) &\le& {1\over (n+1)!}  e^{\b {\mathcal{B}}_v(n+1)} [C_v(\b)]^n\sum_{\tau \in T^0_{n}}1 \\
\nonumber   &=& {(n+1)^{n-1}\over (n+1)!}  e^{\b {\mathcal{B}}_v(n+1)} [C_v(\b)]^n
\end{eqnarray}
where in the last line we have once again used the Cayley formula (see \cite{Ca}). Note that this bound
holds  also when $n=0$, since $1\le e^{\b {\mathcal{B}}_v}$. This concludes the proof of bound \equ(c0n).

~~~~~~~~~~~~~~~~~~~~~~~~~~~~~~~~~~~~~~~~~~~~~~~~~~~~~~~~~~~~~~~~~~~~~~~~~~~~~~~~~~~~~~~~~~~~~~~~~~~~~~~~~~~~~~~~~~~~~~~~~~~~~~~~~~~~~~~~$\Box$
\vskip.2cm

\subsection{Conclusion of the proof of Theorem \ref{main}}
Recalling that we chose $L$ large enough in such a way that $h(L)>R$, the definitions of $\L_h$ and $\L^*_h$ given at the beginning of Section \ref{s3.2} and the Identity \equ(pressure), we can rewrite
the finite volume pressure $\b p^\o_\L(\b,\l)$ of our system as
$$
\b p^\o_\L(\b,\l)= {\l\over |\L|}\left[ \int_{\L_{h}} dx f^\o_\L(x)\Pi^\o_{x,\L}(\b,\l)+ \int_{\L^*_{h}} dx f^\o_\L(x)\Pi^\o_{x,\L}(\b,\l)\right] =
$$
$$
= {\l\over |\L|}\left[ \int_{\L_{h}} dx \Pi^\o_{x,\L}(\b,\l)+ \int_{\L^*_{h}} dx f^\o_\L(x)\Pi^\o_{x,\L}(\b,\l)\right]
$$
where the last identity follows from Remark \ref{f}.

\\By Formula \equ(inside) in Theorem  \ref{t1},  we have that for any $x\in \L_h$
$$
\Pi^\o_{x,\L}(\b,\l) =   P^{\0, n_{h(L)}}_{x,\L}(\b,\l)+Q^{\o,n_{h(L)}}_{x,\L}(\b,\l)
$$
where
$$
P^{\0,n_{h(L)}}_{x,\L}(\b,\l) = \sum_{n=0}^{n_{h(L)}} c^\0_n(x,\b,\L){\l^n}
\Eq(P0)
$$
and
$$
Q^{\o,n_{h(L)}}_{x,\L}(\b,\l)= \sum_{n=n_{h(L)}+1}^{\infty}c^\o_n(x,\b,\L){\l^n}. \Eq(Psio)
$$

\\Therefore, posing
$$
\h^\o_\L(\l,\b)=  {\l\over |\L|}\int_{\L_{h}} dxP^{\0, n_{h(L)}}_{x,\L}(\b,\l)\Eq(eta)
$$
and
$$
\x^\o_\L(\l,\b)= {\l\over |\L|}\left[\int_{\L_{h}} dx Q^{\o,n_{h(L)}}_{x,\L}(\b,\l)+  \int_{\L^*_{h}} dx f^\o_\L(x)\Pi^\o_{x,\L}(\b,\l)\right]\Eq(xi)
$$
we have that $$
\b p^\o_\L(\b,\l)= \h^\o_\L(\l,\b)+ \x^\o_\L(\l,\b).
$$
\\Note that $\h^\o_\L(\l,\b)$ is a polynomial of degree $\l^{n_{h(L)}}$.
\\Let us also define the function
$$
Q^{\0,n_{h(L)}}_{x,\L}(\b,\l)= \sum_{n=n_{h(L)}+1}^{\infty}c^\0_n(x_0,\b,\L){\l^n}. \Eq(Psiob)
$$
\vskip.2cm
\begin{teo}\label{t32}
Let  $\mathcal{D}^\0$ be the closed disc in the complex plane defined in \equ(dv)
$$
\mathcal{D}^\0=\Big\{\l\in \mathbb{C}:~ |\l| \le {1\over e^{\b {\mathcal{B}}_v+1}C_v(\b)}\Big\}.
$$
Then the functions  $\Pi^\0_{x,\L}(\b,\l)$  and   $Q^{\0,n_{h(L)}}_{x,\L}(\b,\l)$ are analytic inside the disc $\mathcal{D}^\0$ where, uniformly in $\L$ and $x$, they admit the bounds
$$
|\Pi^\0_{x,\L}(\b,\l)|\le   (8/7)e^{\b {\mathcal{B}}_v+1}\Eq(bop)
$$
and
$$
|Q^{\0,n_{h(L)}}_{x,\L}(\b,\l)|\le    {e^{\b {\mathcal{B}}_v+1}\over n_{h(L)}^{3/2}}.\Eq(boq)
$$
\end{teo}
\vskip.2cm
{\bf Proof}. We start by proving the analyticity of  $\Pi^\0_{x_0,\L}(\b,\l)$. We have straightforwardly that
$$
|\Pi^\0_{x_0,\L}(\b,\l)|\le
\sum_{n=0}^{\infty} |c^\0_n(x_0,\b,\L)|{|\l|^n},
$$
then, by   Inequality \equ(c0n) in Theorem \ref{t1}, we have
$$
|\Pi^\0_{x_0,\L}(\b,\l)|\le \sum_{n=0}^{\infty}{(n+1)^{n-1}\over (n+1)!}  e^{\b {\mathcal{B}}_v(n+1)} [|\l|C_v(\b)]^n =
e^{\b {\mathcal{B}}_v} \sum_{n=0}^{\infty}{(n+1)^{n-1}\over (n+1)!}  [e^{\b {\mathcal{B}}_v} |\l|C_v(\b)]^n.
$$
Let us now set, for $r\ge 0$
$$
\Th(\b,r)= e^{\b {\mathcal{B}}_v} \sum_{n=0}^{\infty}{(n+1)^{n-1}\over (n+1)!}  [e^{\b {\mathcal{B}}_v} rC_v(\b)]^n
$$
and suppose that there exists
$$
r^*=\max\Big\{r\ge 0: ~~\Th(\b,r)<+\infty \Big\}\Eq(rstar),
$$
then clearly $\Pi^\0_{x_0,\L}(\b,\l)$ is  analytic for all $|\l|\le r^*$ and  $|\Pi^\0_{x_0,\L}(\b,\l)|$ is bounded by $\Th(\b,r^*)$ for all $\l$ in the disc
$|\l|\le r^*$.

To show that this is indeed true, let us just recall the Stirling bound, namely
$$
\sqrt{2\pi n}n^{n}e^{-n}\le n!\le e\sqrt{ n}n^{n}e^{-n} ~~,\Eq(stir)
$$
for all $n\in \mathbb{N}$. So  that we may bound
$$
e^{\b {\mathcal{B}}_v+1} \left[1+{1\over  e}\sum_{n=1}^{\infty}{ [e^{\b {\mathcal{B}_v+1}} rC_v(\b)]^n\over (n+1)^{5\over 2}}\right]\le \Th(\b,r)\le  e^{\b {\mathcal{B}}_v+1} \left[1+{1\over  \sqrt{2\pi}}\sum_{n=1}^{\infty}{ [e^{\b {\mathcal{B}_v+1}} rC_v(\b)]^n\over (n+1)^{5\over 2}}\right].
$$
\vskip.2cm
\\The series $\sum_{n=1}^{\infty}{ (n+1)^{-5/2}[e^{\b {\mathcal{B}_v+1}} rC_v(\b)]^n}$ converges if $e^{\b {\mathcal{B}_v+1}} rC_v(\b)\le 1$ and diverges otherwise. Therefore we get that the number $r^*$ defined in \equ(rstar) does exist and it is equal to
$$
r^*={1\over e^{\b {\mathcal{B}_v+1}} C_v(\b)}.
$$
Moreover,
$$
\Th(\b,r^*)\le  e^{\b {\mathcal{B}}_v+1} \left[1+{1\over  \sqrt{2\pi}}\sum_{n=1}^{\infty}{ 1\over (n+1)^{5\over 2}}\right]< (8/7)e^{\b {\mathcal{B}}_v+1}
$$
and thus we have proved that $\Pi^\0_{x,\L}(\b,\l)$ is analytic in the closed disc $\mathcal{D}^\0$ and its modulus is bounded   there by $(8/7)e^{\b {\mathcal{B}}_v+1}$.

Let us now prove the analyticity and boundedness of  the function $Q^{\0,n_{h(L)}}_{x,\L}(\b,\l)$ defined in \equ(Psiob) when $\l$ varies in the complex disc $\mathcal{D}^\0$. Using once again inequality \equ(c0n)  and the Stirling bound \equ(stir)  and assuming that $|\l|$ varies in $\mathcal{D}^\0$,
we have
$$
|Q^{\0,n_{h(L)}}_{x,\L}(\b,\l)|= \sum_{n=n_{h(L)}+1}^{\infty}|c^\0_n(x_0,\b,\L)|{|\l|^n}\le  \sum_{n=n_{h(L)}+1}^{\infty}
{(n+1)^{n-1}\over (n+1)!}  e^{\b {\mathcal{B}}_v(n+1)} [C_v(\b)]^n{|\l|^n} \le
$$
$$
\le {e^{\b {\mathcal{B}}_v+1}\over  \sqrt{2\pi}}\sum_{n=n_{h(L)}+1}^{\infty}{[e^{\b {\mathcal{B}}_v+1} C_v(\b)|\l|]^n\over (n+1)^{5\over 2}}
\le {e^{\b {\mathcal{B}}_v+1}\over  \sqrt{2\pi}}\sum_{n=n_{h(L)}+1}^{\infty}{1\over (n+1)^{5\over 2}} \le {e^{\b {\mathcal{B}}_v+1}\over  \sqrt{2\pi}}\sum_{n=n_{h(L)}+1}^{\infty}{1\over n^{5/2}}\le
$$
$$
\le {e^{\b {\mathcal{B}}_v+1}\over  \sqrt{2\pi}}\int_{n_{h(L)}}^\infty{1\over u^{5/2}}du=
{2e^{\b {\mathcal{B}}_v+1}\over  3\sqrt{2\pi}} {1\over n_{h(L)}^{3/2}}<    {e^{\b {\mathcal{B}}_v+1}\over n_{h(L)}^{3/2}}.
$$
Hence we have proved that $|Q^{\0,n_{h(L)}}_{x,\L}(\b,\l)|$ is analytic in $\mathcal{D}^\0$ and bounded there according to \equ(boq).

\vskip.2cm
\begin{teo}\label{t33}
Given $x\in \L$ and $\o\in \O^*$,
let  $\mathcal{D}^\o$ the closed disc in the complex plane defined in \equ(dnv)
$$
\mathcal{D}^\o=\Big\{\l\in \mathbb{C}:~ |\l| \le {1\over e^{\k\b\r_\o}e^{\b \mathcal{B}_v+1}C_v(\b)}\Big\}.
$$
Then the function $\Pi^{\o}_{x,\L}(\b,\l)$ defined in \equ(1.3) is analytic in the disc $\mathcal{D}^\o$ where, uniformly in $\L$, $x$ and $\o$
$$
|\Pi^{\o}_{x,\L}(\b,\l)|\le (8/7) {e^{\b {\mathcal{B}}_v+1}}.\Eq(bp)
$$
Moreover, let $n_{h(\L)}$ be the integer defined in \equ(nl), then the function $Q^{\o,n_{h(L)}}_{x,\L}(\b,\l)$ defined in \equ(Psio) is analytic in the disc $\mathcal{D}^\o$ where, uniformly in $\L$, $x$ and $\o$
$$
|Q^{\o,n_{h(L)}}_{x,\L}(\b,\l)|\le {e^{\b {\mathcal{B}}_v+1}\over n_{h(L)}^{3/2}}.\Eq(bq)
$$
\end{teo}
\vskip.2cm
{\bf Proof}. \\ The proof of \equ(bp) and \equ(bq) proceeds along the same lines described in the previous theorem. In order to  proof bound \equ(bp), we can use the bound \equ(cogen) on $ |c^\o_n(x_0,\b,\L)|$ given in Proposition \ref{Cgene}  together with the Stirling bound \equ(stir)  to get, for all $\l\in \mathcal{D}^\o$,
$$
|\Pi^{\o}_{x,\L}(\b,\l)|= \sum_{n=0}^{\infty}|c^\o_n(x_0,\b,\L)|{|\l|^n}\le  \sum_{n=0}^{\infty}
{(n+1)^{n-1}\over (n+1)!}  e^{\b {\mathcal{B}}_v(n+1)} [C_v(\b)e^{\k\b\r_\o}]^n{|\l|^n} \le
$$
$$
\le e^{\b {\mathcal{B}}_v+1}\left[1+{1\over  \sqrt{2\pi}}\sum_{n=0}^{\infty}{[e^{\b {\mathcal{B}}_v+1}e^{\k\b\r_\o} C_v(\b)|\l|]^n\over (n+1)^{5\over 2}}\right]\le
e^{\b {\mathcal{B}}_v+1}\left[1+{1\over  \sqrt{2\pi}}\sum_{n=0}^{\infty}{1\over (n+1)^{5\over 2}}\right]\le (8/7)e^{\b {\mathcal{B}}_v+1}.
$$

\vskip.2cm

Concerning now the bound \equ(bq) we have,
$$
|Q^{\o,n_{h(L)}}_{x,\L}(\b,\l)|= \sum_{n=n_{h(L)}+1}^{\infty}|c^\o_n(x_0,\b,\L)|{|\l|^n}\le  \sum_{n=n_{h(L)}+1}^{\infty}
{(n+1)^{n-1}\over (n+1)!}  e^{\b {\mathcal{B}}_v(n+1)} [C_v(\b)e^{\k\b\r_\o}]^n{|\l|^n} \le
$$
$$
\le {e^{\b {\mathcal{B}}_v+1}\over  \sqrt{2\pi}}\sum_{n=n_{h(L)}+1}^{\infty}{[e^{\b {\mathcal{B}}_v+1}e^{\k\b\r_\o} C_v(\b)|\l|]^n\over (n+1)^{5\over 2}}
\le {e^{\b {\mathcal{B}}_v+1}\over  \sqrt{2\pi}}\sum_{n=n_{h(L)}+1}^{\infty}{1\over (n+1)^{5\over 2}} \le
$$
$$
\le {e^{\b {\mathcal{B}}_v+1}\over  \sqrt{2\pi}}\sum_{n=n_{h(L)}+1}^{\infty}{1\over n^{5/2}}\le {e^{\b {\mathcal{B}}_v+1}\over  \sqrt{2\pi}}\int_{n_{h(L)}}^\infty{1\over u^{5/2}}du=
{2e^{\b {\mathcal{B}}_v+1}\over  3\sqrt{2\pi}} {1\over n_{h(L)}^{3/2}}<    {e^{\b {\mathcal{B}}_v+1}\over n_{h(L)}^{3/2}}.
$$

\vskip.2cm
\begin{pro} The function $\h^\o_\L(\l,\b)$ defined in \equ(eta) is analytic in the whole complex plane  and its modulus is bounded as
$$
|\h^\o_\L(\l,\b)|\le  (8/7)e^{\b {\mathcal{B}}_v+1}|\l| \Eq(bhl)
$$
as $\l$ varies in the disc $\mathcal{D}^\0$. Moreover  for $\l\in\mathcal{D}^\0(\b)$ it holds that
$$
\lim_{\L\to\infty} \h^\o_\L(\l,\b)=\b p^\0(\b,\l)\Eq(limh)
$$
\end{pro}
\vskip.2cm
\\{\bf Proof}.  The analyticity of $\h^\o_\L(\l,\b)$ in the whole complex plane   follows trivially from the fact that, by definition \equ(eta),  $\h^\o_\L(\l,\b)$ is, as a function of $\l$,  a polynomial of degree $n_{h(L)}$. The fact that the modulus $|\h^\o_\L(\l,\b)|$ is bounded by  the r.h.s. of \equ(bhl) when $\l\in\mathcal{D}^\0$ follows trivially from bound  \equ(bop) of Theorem \ref{t32}. Indeed, if $\l\in\mathcal{D}^\0$,
recalling the definition \equ(P0) of $P^{\0, n_{h(L)}}_{x,\L}(\b,\l)$, it easily follows from Theorem \ref{t32} (see the proof of inequality \equ(bop))  that
$$
|P^{\0, n_{h(L)}}_{x,\L}(\b,\l)|\le \sum_{n=0}^{n_{h(L)}} |c^\0_n(x_0,\b,\L)|{|\l|^n}\le
\sum_{n=0}^{\infty} |c^\0_n(x_0,\b,\L)|{|\l|^n}\le (8/7)e^{\b {\mathcal{B}}_v+1}. \Eq(Pbo)
$$
Therefore, if $\l\in\mathcal{D}^\0$,
$$
|\h^\o_\L(\l,\b)|\le  {|\l|\over |\L|}\int_{\L_{h}} dx|P^{\0, n_{h(L)}}_{x,\L}(\b,\l)|\le  {\l\over |\L|}\int_{\L_{h}}  (8/7)e^{\b {\mathcal{B}}_v+1}dx\le  |\l| (8/7)e^{\b {\mathcal{B}}_v+1}.
$$

In order to prove \equ(limh), observe that
$$
\b p^\0_\L(\b,\l)~=~{\l\over |\L|}\int_\L dx \Pi^\0_{x_0,\L}(\b,\l) =  ~{\l\over |\L|}\int_\L dx\left(P^{\0, n_{h(L)}}_{x,\L}(\b,\l)+Q^{\0, n_{h(L)}}_{x,\L}(\b,\l)\right),
$$
so that
$$
\h^\o_\L(\l,\b)- \b p^\0_\L(\b,\l)=  {\l\over |\L|}\Bigg[\int_{\L_{h}} dxP^{\0, n_{h(L)}}_{x,\L}(\b,\l)- \int_\L dxP^{\0, n_{h(L)}}_{x,\L}(\b,\l)
-\int_{\L} dxQ^{\0, n_{h(L)}}_{x,\L}(\b,\l)\Bigg],
$$
hence
$$
|\h^\o_\L(\l,\b)- \b p^\0_\L(\b,\l)|\le  {|\l|\over |\L|}\left[\int_{\L\setminus\L_{h}} dx|P^{\0, n_{h(L)}}_{x,\L}(\b,\l)|+\int_{\L} dx |Q^{\0, n_{h(L)}}_{x,\L}(\b,\l)|\right].
$$

Now, if $\l\in\mathcal{D}^\0$,  by \equ(Pbo)  and \equ(boq) we have that
$$
|P^{\0, n_{h(L)}}_{x,\L}(\b,\l)|\le (8/7)e^{\b {\mathcal{B}}_v+1}
$$
and
$$
|Q^{\0, n_{h(L)}}_{x,\L}(\b,\l)|\le  {e^{\b {\mathcal{B}}_v+1}\over n_{h(L)}^{3/2}}
$$
Hence
$$
|\h^\o_\L(\l,\b)- \b p^\0_\L(\b,\l)|\le |\l|e^{\b {\mathcal{B}}_v+1}\left[ (8/7) {|\L\setminus \L_h|\over |\L|}+{1\over n_{h(L)}^{3/2}}\right]=
|\l|e^{\b {\mathcal{B}}_v+1}\left[ (8/7) {|\L^*_h|\over |\L|}+{1\over n_{h(L)}^{3/2}}\right].
$$
Recalling \equ(due), we have that for $\l\in\mathcal{D}^\0$
$$
\lim_{\L\to\infty} |\h^\o_\L(\l,\b)- \b p^\0_\L(\b,\l)|\le \lim_{\L\to\infty} |\l|e^{\b {\mathcal{B}}_v+1}\left[ (8/7) {|\L^*_h|\over |\L|}+{1\over n_{h(L)}^{3/2}}\right]=0\Eq(ofco)
$$
where in the last line we used  \equ(due) and \equ(limn).

Of course \equ(ofco) implies that, for  all $\l\in\mathcal{D}^\0$, it holds that
$$
\lim_{\L\to\infty} \h^\o_\L(\l,\b)= \lim_{\L\to\infty}  \b p^\0_\L(\b,\l)= \b p^\0(\b,\l),
$$
which ends the proof.

~~~~~~~~~~~~~~~~~~~~~~~~~~~~~~~~~~~~~~~~~~~~~~~~~~~~~~~~~~~~~~~~~~~~~~~~~~~~~~~~~~~~~~~~~~~~~~~~~~~~~~~~~~~~~~~~~~~~~~~~~~~~~~~~$\Box$

\vskip.2cm

\begin{pro}\label{profin}
The function $\x^\o_\L(\l,\b)$ defined in  \equ(xi)   is analytic and bounded as far as $\l\in \mathcal{D}^\o$.
Moreover, it holds that
$$
\lim_{\L\to\infty}\x^\o_\L(\l,\b)=0.\Eq(limu)
$$
\end{pro}
\vskip.2cm

\\{\bf Proof.} We recall the definition \equ(xi) of $\x^\o_\L(\l,\b)$
$$
\x^\o_\L(\l,\b)={\l\over |\L|}\left[\int_{\L_{h}} dx Q^{\o,n_{h(L)}}_{x,\L}(\b,\l)+  \int_{\L^*_{h}} dx f^\o_\L(x)\Pi^\o_{x,\L}(\b,\l)\right].
$$
By Theorem \ref{t33},  both $Q^{\o,n_{h(L)}}_{x,\L}(\b,\l)$ and $\Pi^\o_{x,\L}(\b,\l)$ are analytic and bounded in the closed disc $\mathcal{D}^\o$.
This immediately implies that $\x^\o_\L(\l,\b)$ is also analytic (and bounded) in $\mathcal{D}^\o$.

\\Concerning the limit \equ(limu), we
can use the bounds \equ(bp) and \equ(bq)   given in Theorem \ref{t33} for $\Pi^\o_{x,\L}(\b,\l)$ and $Q^{\o,n_{h(L)}}_{x,\L}(\b,\l)$  respectively and, recalling Remark \ref{f}, we have that
$f^\o_\L(x)\le e^{\b\k\r_\o}$ and so
\begin{eqnarray}
\nonumber |\x^\o_\L(\l,\b)| &\le&{|\l|\over |\L|}\left[\int_{\L_{h}} dx |Q^{\o,n_{h(L)}}_{x,\L}(\b,\l)|+  \int_{\L^*_{h}} dx f^\o_\L(x)|\Pi^\o_{x,\L}(\b,\l)|\right]
\\
\nonumber &\le& {|\l| \over |\L|}\left[\int_{\L_{h}} dx {e^{\b {\mathcal{B}}_v+1}\over n_{h(L)}^{3/2}}+  \int_{\L^*_{h}} dx e^{\b\k\r_\o}(8/7) {e^{\b {\mathcal{B}}_v+1}}\right]\\
\nonumber&\le&  (8/7)|\l| e^{\b\k\r_\o} e^{\b {\mathcal{B}}_v+1} \Bigg[{|\L_h|\over |\L| n_{h(L)}^{3/2} }+  {|\L^*_h|\over |\L|}\Bigg].
\end{eqnarray}

\\Hence
\begin{eqnarray}
\nonumber\lim_{\L\to\infty}|\x^\o_\L(\l,\b)| & \le & (8/7)|\l| e^{\b\k\r_\o} e^{\b {\mathcal{B}}_v+1}  \lim_{\L\to\infty} \Bigg[{|\L_h|\over |\L| n_{h(L)}^{3/2} }+  {|\L^*_h|\over |\L|}\Bigg]\\
\nonumber &=& (8/7) |\l| e^{\b\k\r_\o} e^{\b {\mathcal{B}}_v+1} \Bigg[ \lim_{\L\to\infty}{|\L_h|\over |\L|}  \lim_{\L\to\infty}{1\over  n_{h(L)}^{3/2}} +   \lim_{\L\to\infty}{|\L^*_h|\over |\L|}\Bigg]\\
\nonumber &=&  0
\end{eqnarray}
where in the last line, by the definitions given at the beginning of Secion \ref{s3.2}, we have used that
$$
\lim_{\L\to\infty}{|\L_h|\over |\L|}=1~,~~~~~~~\lim_{\L\to\infty}n_{h(L)}=+\infty~,~~~~~~~~ \lim_{\L\to\infty}{|\L^*_h|\over |\L|}=0.
$$
This ends the proof of Proposition \ref{profin}.

~~~~~~~~~~~~~~~~~~~~~~~~~~~~~~~~~~~~~~~~~~~~~~~~~~~~~~~~~~~~~~~~~~~~~~~~~~~~~~~~~~~~~~~~~~~~~~~~~~~~~~~~~~~~~~~~~~~~~~~~~~~~~~~~~~~~~~~~~~~~~~~~~~~~~~$\Box$

\end{document}